\documentclass[conference]{IEEEtran}

\usepackage{cite}
\usepackage{graphicx}
\usepackage{epstopdf}
\usepackage{amsmath}
\usepackage{amsfonts}
\usepackage{algorithmic}
\usepackage{array}
\usepackage{stfloats}
\usepackage{multirow}
\usepackage{color}

\DeclareMathOperator*{\argmax}{arg\,max}

\begin{document}

\title{Efficiency Analysis of Decoupled Downlink and Uplink Access in Heterogeneous Networks}

\author{\IEEEauthorblockN{Katerina Smiljkovikj}
\IEEEauthorblockA{Ss. Cyril and Methoduius University\\
Skopje, Macedonia\\
Email: katerina@feit.ukim.edu.mk}
\and
\IEEEauthorblockN{Liljana Gavrilovska}
\IEEEauthorblockA{Ss. Cyril and Methoduius University\\
Skopje, Macedonia\\
Email: liljana@feit.ukim.edu.mk}
\and
\IEEEauthorblockN{Petar Popovski}
\IEEEauthorblockA{Aalborg University\\
Aalborg, Denmark\\
Email: petarp@es.aau.dk}}


%

\maketitle

\begin{abstract}
This paper analyzes two-tier heterogeneous cellular network with decoupled downlink and uplink access. The basic performance benefits of uplink/downlink decoupling have been recently introduced. Here we provide a more elaborate treatment of the decoupling mechanism by analyzing spectral and energy efficiency of the system, as the joint improvement of these two features is crucial for the upcoming 5G systems. Contrary to the common assumption of a homogeneous user domain, we analyze two-tier user domain, whose transmit powers depend purely on the association process.
The derived joint association probabilities and the distributions of the distance to the serving base station give deeper insight into the fundamentals of a system with decoupled access. The rigorous theoretical analysis shows that decoupling of downlink and uplink with two-level uplink power adaptation improves both, spectral and energy efficiency of the system. 
\end{abstract}



%
\IEEEpeerreviewmaketitle

\section{Introduction}

The evolution of cellular networks is constantly progressing towards increased heterogeneity of network infrastructure, in order to respond to the ever increasing traffic demands. From homogeneous cellular networks consisting of macro base stations only to heterogeneous cellular networks consisting of multiple types of base stations, the evolution of the infrastructure offers plenty of challenges for network engineers \cite{GC_4}. Heading towards the fifth generation (5G), there is a broad consensus that one of the major requirements of future networks is higher spectral efficiency and higher energy efficiency \cite{SE_EE,5G}.

The increasing heterogeneity requires reconsidering some fundamental mechanisms in the networks, one of which is the cell association process. For some devices in the network, with multiple base stations in the surrounding, an optimal base station for downlink access does not necessarily need to be optimal for the uplink access. One of the main reasons is the difference in the signal and interference power in both directions. Therefore, in certain circumstances and for certain devices, it is beneficial to decouple downlink and uplink and treat them as two independent communication links, which leads to the concept of \emph{decoupled downlink/uplink access}.

The decoupled access was first indicated in \cite{GC_2}, where the author suggests decoupling due to the difference in signal power and interference in both directions. Furthermore, the authors in \cite{SE_EE} suggest decoupling in order to achieve fair allocation between the resources from different base stations, i.e. it can be beneficial for a device to associate to less loaded base station with worse channel conditions than to highly loaded base station with better channel conditions. The most recent work is the one conducted in \cite{DUDe,KaterinaPetarLiljanaDUD}. The authors in \cite{DUDe} show the throughput benefits using real-world simulator, while the authors in \cite{KaterinaPetarLiljanaDUD} derive the joint association probabilities for two-tier network with decoupled access and show the improvement in the uplink coverage probability. The decoupled access was also mentioned in \cite{TonyD2D} in terms of Device-to-Device (D2D) communications, where the authors assume that downlink and uplink are two separate links and perform the analysis separately. 

In this paper we present theoretical analysis of a two-tier heterogeneous cellular system with decoupled downlink (DL) and uplink (UL) access, in dense user environment. We introduce simple two-level uplink power adaptation and evaluate the system performance for varying uplink power levels. The system model is based on stochastic geometry, which is successfully used for cellular networks modeling \cite{Baccelli}. The system is also suitable for machine-type devices that have a much higher density compared to the density of the base stations. Additional feature of machine-type traffic is their high uplink/downlink ratio, which gains in significance with the improved uplink coverage with decoupled access \cite{KaterinaPetarLiljanaDUD}. The results presented in this paper reveal that, under certain circumstances, we can achieve improvement, both for the spectral efficiency and the energy efficiency in the system with decoupled DL and UL, only by using a simple, two-level uplink power adaptation.

The paper is organized as follows. Section \ref{SysModel} describes the system model used to evaluate the decoupled access. Section \ref{sec:Analysis} contains the main results of the association probability, the distribution of the distance to the serving base station and spectral and energy efficiency. Section \ref{sec:NumericalResults} give the numerical results regarding the analysis and Section \ref{conclusion} concludes the paper.




 

\section{System model}
\label{SysModel}

We model a two-tier heterogeneous cellular network, consisting of Macro Cell (Mcell) tier and Small Cell (Scell) tier and heterogeneous user domain. The nodes in the system are deployed by homogeneous Poisson Point Process (PPP) $\Phi_v$ with intensity measure $\lambda_v$. A point in $\mathbb{R}^2$, that is a realization of a PPP $\Phi_v$ is denoted as $x_v$. For Mcell Base Stations (MBSs), $v=M$, for Scell Base Stations (SBSs), $v=S$ and for the devices $v=d$. The transmit power of MBSs is $P_M$ and the transmit power of SBSs is $P_S$, such that $P_S<P_M$. Assuming $\lambda_d>>(\lambda_M+\lambda_S)$, there is always one active device per base station on a dedicated resource unit. The transmit power of the devices associated to MBSs and SBSs is $Q_M$ and $Q_S$, respectively. While for the DL transmissions it is per definition clear that $P_S>P_M$, we are not predefining a similar relationship for $Q_M$ and $Q_S$. However, it is intuitively expected that $Q_M>Q_S$, because more power is needed to transmit to a more distant base stations. For homogeneous user domain, $Q_M=Q_S$. On the other hand, we make the following assumption about the ratio of the transmit powers: 
\begin{equation} \label{RatioQSQMPSPM}
\frac{Q_S}{Q_M} \geq \frac{P_S}{P_M}
\end{equation}
The rationale for this will become clear in relation to the association regions on Fig.~\ref{fig:AssocRegions}. The analysis is performed on a typical device, located at the origin, which is allowed by Slivnyak's theorem \cite{Chiu}.

The received signal in DL at the typical device from a base station located at $x_v$ and the received signal in UL at a base station located at $x_v$ are given by:
\begin{eqnarray}
\textrm{S}_{DL,v} &=& P_v h_{x_v} \left\|x_v\right\|^{-\alpha} \label{DLsignal} \\
\textrm{S}_{UL,v} &=& Q_v h_{x_v} \left\|x_v\right\|^{-\alpha} \label{ULsignal}
\end{eqnarray}  
\noindent where $v \in \{M,S\}$ and consequently $x_v$ can be a point from $\Phi_M$ or $\Phi_S$. $h_{x_v}$ describes the fast fading process at the point $x_v$. We assume Rayleigh fading with independent identically distributed (i.i.d) magnitude with unit mean $\left(h\sim \textrm{exp}(1)\right)$. Due to the i.i.d. assumption, we will omit the subscript whenever there is no danger to cause confusion. 

A device applies two separate rules for association in the DL and the UL, respectively. The association is based on maximum average received signal power, averaged over fast fading. The following equations formulate the DL and the UL association rule, respectively:
\begin{eqnarray}
i = \underset{v \in \{M,S\}}\argmax{\mathbb{E}_{h}\left[ S_{DL,v} \right]} &=& \underset{v \in \{M,S\}}\argmax{P_v \left\|x_v\right\|^{-\alpha}} \label{DLsignal_average} \\
j = \underset{v \in \{M,S\}}\argmax{\mathbb{E}_{h}\left[ S_{UL,v} \right]} &=& \underset{v \in \{M,S\}}\argmax{Q_v \left\|x_v\right\|^{-\alpha}} \label{ULsignal_average}
\end{eqnarray}    
The concept of decoupled access implies that $i$ can be different than $j$ and the device can be associated to two different base stations, each one associated with one direction, DL or UL. 

The analysis is focused on the UL. Basically, by decoupling DL and UL and adapting the UL association to the signals transmitted in UL, there is an improved Signal-to-Interference-plus-Noise-Ratio (SINR) for the UL \cite{KaterinaPetarLiljanaDUD}. The typical device has interference from devices associated to other base stations, transmitting on the same resource unit. Under the assumption of orthogonal resource allocation in one cell, we model the number of interfering devices to be equal to the number of base stations (one interferer per base station). The interfering devices are modeled by independent homogeneous PPP $\Phi_{I_d}$ with intensity measure $\lambda_{I_d}$, such that $\lambda_{I_d}=\lambda_M+\lambda_S$. Alternative way to define the spatial process for the interfering machines is by thinning \cite{Chiu} the original point process $\Phi_d$ with probability $p=\frac{\lambda_M+\lambda_S}{\lambda_d}$. Although this model of interference is an approximation, due to the dependence introduced by the association process, the authors in \cite{NovlanULcellNet} show that this dependence is weak and it is justified to approximate the interfering devices by independent homogeneous PPP. 

We assume that each transmission from a device requires one resource unit. Putting this in the context of LTE, the resource unit is the Resource Block (RB), which is the minimum scheduling unit that can be assigned to a particular device. In the context of M2M traffic, this can be interpreted as having already reserved resources for the machines and each machine is assigned one resource unit for one transmission. The main purpose of this paper is to evaluate the performance of decoupled access with the simple power adaptation, based purely on the signal power improvement in UL along with the association process. For resource reservation with high reliability, one can refer to the schemes presented in \cite{PetarM2Mreporting}.

The presented system is evaluated in terms of spectral and energy efficiency. The spectral efficiency $S_{eff}$ is defined as the channel capacity normalized by the system bandwidth. The energy efficiency $E_{eff}$ is defined as the channel capacity normalized by the system power consumption. These quantities are analytically defined as~\cite{SE_EE}:
\begin{eqnarray}
	S_{eff} &=& \textrm{log}_2 \left( 1+SINR\right) \label{SEdef} \\
	E_{eff} &=& \frac{W}{P_{tot}}\textrm{log}_2 \left( 1+SINR\right) \label{EEdef}
\end{eqnarray}
\noindent where $W$ is the allocated frequency bandwidth. The total power consumption ($P_{tot}$) is defined as \cite{SE_EE}:
\begin{equation}
P_{tot} = \frac{Q_v}{\rho}+P_C
\end{equation}
\noindent where $v \in \{M, S\}$, $\rho$ is the power amplifier efficiency and $P_C$ is the circuit consumption. We will use the general definitions in (\ref{SEdef}) and (\ref{EEdef}) to derive the spectral and energy efficiency for the system described above. Our main purpose is to analyze the impact that the simple power adaptation has on these two quantities, $S_{eff}$ and $E_{eff}$. 

\section{Analysis}
\label{sec:Analysis}

In this section, we first derive the joint association probability and the distance distributions to the serving base stations. We then use these results to derive the spectral and energy efficiency.

\subsection{Association probability}
\label{sec:AssocProb}

The association process can lead to one of the four possible association cases: (i) Case 1: DLAP=ULAP=MBS; (ii) Case 2: DLAP=MBS and ULAP=SBS; (iii) Case 3: DLAP=SBS and ULAP=MBS  and (iv) Case 4: DLAP=ULAP=SBS. The joint association probabilities for homogeneous user domain (all devices transmit with the same transmit power) are already elaborated in \cite{KaterinaPetarLiljanaDUD}. We follow the same procedure and derive the association probabilities for the model described in this paper, where devices transmit with two power levels, depending on the type of base station they are transmitting to. 

\begin{figure}[t!]
	\centering
		\includegraphics[width=8.3cm]{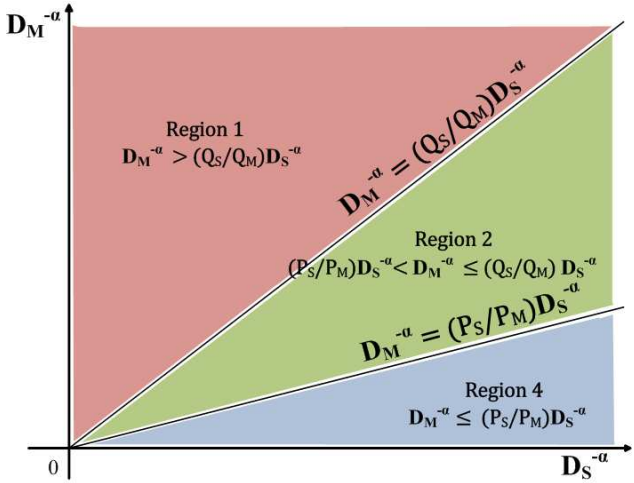}
		\caption{Association regions for the four possible cases of association in two-tier heterogeneous network.}
		\label{fig:AssocRegions}
\end{figure}

Let $D_M$ and $D_S$ denote the distance from the nearest MBS and SBS to the typical point at the origin, which probability density function (pdf) and cumulative distribution function (cdf) are derived using contact distributions of a point process (the probability that there in no point in the circle with radius $x$ \cite{Chiu}) and are given by:
\begin{eqnarray}
	f_{D_v}(x) &=& 2\pi{\lambda}_v x e^{-\pi{\lambda}_v x^2}, x\geq0 \label{contactPDF} \\
	F_{D_v}(x) &=& 1-e^{-\pi{\lambda}_v x^2}, x\geq0 \label{contactCDF}
\end{eqnarray}
Rephrasing the association rules given by (\ref{DLsignal_average}) and (\ref{ULsignal_average}), one can say that a device is associated to MBS in DL if $P_MD_M^{-\alpha}>P_SD_S^{-\alpha}$ and it is associated to MBS in UL if $Q_MD_M^{-\alpha}>Q_SD_S^{-\alpha}$. Each association case is defined by an association region, Fig.~\ref{fig:AssocRegions}, and association probability. 

Following the procedure described in \cite{KaterinaPetarLiljanaDUD}, the association probabilities are derived as:
\subsubsection{Case 1: DLAP=ULAP=Mcell}
\begin{flalign}
		\Pr(\textrm{Case }1) &= \frac{{\lambda}_M}{{\lambda}_M + \left(\frac{Q_S}{Q_M}\right)^{2/\alpha}{\lambda}_S}&&
		\label{Pcase1}
\end{flalign}

\subsubsection{Case 2: DLAP=Mcell\&ULAP=Scell}
\begin{flalign}
		\Pr (\textrm{Case }2) &= \frac{{\lambda}_M}{{\lambda}_M + \left(\frac{P_S}{P_M}\right)^{2/\alpha}{\lambda}_S} - \frac{{\lambda}_M}{{\lambda}_M + \left(\frac{Q_S}{Q_M}\right)^{2/\alpha}{\lambda}_S}&&
		\label{Pcase2}
	\end{flalign}

Here we note that the condition (\ref{RatioQSQMPSPM}) ensures that $\Pr (\textrm{Case }2) \geq 0$. Otherwise, if the condition (\ref{RatioQSQMPSPM}) is not satisfied, then the probability of Case 2 is strictly zero.

\subsubsection{Case 3: DLAP=Scell\&ULAP=Mcell}
\begin{flalign}
	\Pr (\textrm{Case }3) &= 0&&
\end{flalign}

\subsubsection{Case 4: DLAP=ULAP=Scell}
\begin{flalign}
	\Pr (\textrm{Case }4) &= \frac{{\lambda}_M}{{\lambda}_M + \left(\frac{P_S}{P_M}\right)^{2/\alpha}{\lambda}_S}&&
		\label{Pcase4}
\end{flalign}
 
Varying the UL transmit powers $Q_M$ and $Q_S$, we can change the probabilities for Cases 1 and 2, i.e. the probability for Case 1 can increase at the expense of the probability for Case 2 and vise verse. Varying the DL transmit powers $P_M$ and $P_S$ we can make trade-offs between the probabilities for Cases 2 and 4.

\subsection{Distance distribution}
\label{sec:DistanceDist}

By decoupling DL/UL, we affect the association process, but also the distance to the base station to which the device is associated, since the device chooses a base station that is closer, compared to the association that does not consider decoupling. In this section, we derive the pdf of the distance to the serving base station, as a direct consequence of the association process. The association probabilities given in Section \ref{sec:AssocProb} state that a particular device can belong to one of the three regions, 1, 2 and 4. The probability for Case 3 is zero and none of the devices will be associated with this combination of base stations. Let $D_{v|i}$ denote the distance to the serving base station conditioned on Case $i$, i.e. the distance to the serving base station for the devices that are located in Region $i$, where $i \in \{1,2,4\}$. The pdf of the distance to the serving base station is derived for the three feasible regions. 

\subsubsection{Case 1}
The devices in Region 1 are associated to MBS in UL and the distance to their serving base station satisfies $D_{M}^{-\alpha}>\frac{Q_S}{Q_M}D_{S}^{-\alpha}$. The complementary cumulative distribution function (ccdf) is calculated as:
\begin{align}
		&F_{D_{M|1}}^c (x) = \Pr \left( D_{M}>x \mid D_{S}^{-\alpha}<\frac{Q_M}{Q_S}D_{M}^{-\alpha} \right) \nonumber \\
		&= \frac{\Pr \left( D_M > x ; D_M < \left(\frac{Q_M}{Q_S}\right)^{1/\alpha}D_S \right)}{\Pr(\textrm{Case }1)} \nonumber \\
		&= \frac{\int\limits_{x}^{\infty} \left(1 - e^{-\pi {\lambda}_M \left(\frac{Q_M}{Q_S}\right)^{2/\alpha} x_s^2} \right) f_{D_S}(x_s) \mathrm{d}{x_s} } {\Pr(\textrm{Case }1)}
		\label{CondCDF}
	\end{align}
where $\Pr(\textrm{Case }1)$ is given by~(\ref{Pcase1}) and $f_{D_S}(x_s)$ is given by (\ref{contactPDF}) for $v=S$. The cdf of the distance is $F_{D_{M|1}}(x) = 1 - F_{D_{M|1}}^c(x)$. By differentiating the cdf, we derive the pdf of the distance to the serving base station for the devices in Region 1:
\begin{eqnarray}
		f_{D_{M|1}} (x) &=& \frac{\left(1 - e^{-\pi {\lambda}_M \left(\frac{Q_M}{Q_S}\right)^{2/\alpha} x^2} \right) f_{D_S}(x)} {\Pr(\textrm{Case }1)}
		\label{CondPDF_DUD}
	\end{eqnarray}

\subsubsection{Case 2: DLAP=MBS\&ULAP=SBS}

The devices associated with Case 2 are associated to SBS in UL and the distance to their serving base station satisfies $\frac{P_S}{P_M}D_{S}^{-\alpha} < D_{M}^{-\alpha} \leq \frac{Q_S}{Q_M}D_{S}^{-\alpha}$. The pdf of the distance to the serving base station is given by:

\begin{eqnarray}
		f_{D_{S|2}} (x) = \left(e^{-\pi {\lambda}_S \left(\frac{P_S}{P_M}\right)^{\frac{2}{\alpha}} x^2} - e^{-\pi {\lambda}_S \left(\frac{Q_S}{Q_M}\right)^{\frac{2}{\alpha}} x^2} \right) \times \nonumber \\
\frac{f_{D_M}(x)} {\Pr(\textrm{Case }2)}
		\label{CondPDF_DUD}
	\end{eqnarray}
\noindent where $\Pr(\textrm{Case }2)$ is given by (\ref{Pcase2}) and $f_{D_M}(x)$ is given by (\ref{contactPDF}) for $v=M$.

\subsubsection{Case 4: DLAP=ULAP=SBS} 

The devices associated with Case 4 are associated to SBS in UL and the distance to their serving base station satisfies $D_{S}^{-\alpha} \geq \frac{P_M}{P_S}D_{M}^{-\alpha}$. The pdf of the distance to the serving base station is given by:

\begin{eqnarray}
		f_{D_{S|4}} (x) = \frac{\left(e^{-\pi {\lambda}_M \left(\frac{P_M}{P_S}\right)^{2/\alpha} x^2} \right) f_{D_S}(x)} {\Pr(\textrm{Case }4)}
		\label{CondPDF_DUD}
\end{eqnarray}
\noindent where $\Pr(\textrm{Case }4)$ is given by (\ref{Pcase4}) and $f_{D_S}(x)$ is given by (\ref{contactPDF}) for $v=S$.

\subsection{Spectral and energy efficiency}

In this section, we derive the results for spectral efficiency, while the results for energy efficiency rely on the same type of derivation. Given the association probabilities in Section \ref{sec:AssocProb} and the distance distributions in Section \ref{sec:DistanceDist}, the average spectral efficiency in the system can be calculated as:
\begin{eqnarray}
		S_{eff} = \sum_{i=1}^4 {S_{eff}(\textrm{Case }i)\Pr(\textrm{Case }i)}
		\label{SEdef_average}
\end{eqnarray}
\noindent where $S_{eff}(\textrm{Case }i)$ is the spectral efficiency of the devices in Region $i$. The interfering devices transmit with two power levels, depending on the base station they are associated to. In this model, $\frac{\lambda_M}{\lambda_M+\lambda_S}$ percent are transmitting with power $Q_M$ and $\frac{\lambda_S}{\lambda_M+\lambda_S}$ percent are transmitting with power $Q_S$. Using the approach derived in \cite{KaterinaPetarLiljanaDUD}, we define new discrete random variable $Z$, which has two values, $Q_M$ and $Q_S$ with probabilities that are proportional to their densities:
\begin{eqnarray}
\Pr(Q_M) &=& \frac{\lambda_M}{\lambda_M+\lambda_S} \\
\Pr(Q_S) &=& \frac{\lambda_S}{\lambda_M+\lambda_S}
\end{eqnarray}
The UL signal originating from an interfering device is: 
\begin{eqnarray}
S_{U} = Z h \left\|x_{I_m}\right\|^{-\alpha} = h \left\|Z^{-\frac{1}{\alpha}} x_{I_m}\right\|^{-\alpha} = h \left\|y_{I_m}\right\|^{-\alpha}
\label{abc}
\end{eqnarray}
\noindent where $x_{I_m}$ are points from $\Phi_{I_m}$. We use $y_{I_m}$ to denote the spatial point of an equivalent PPP $\widetilde{\Phi}_{I_m}$, which represents the points from $\Phi_{I_m}$, but randomly dislocated through the value of the power random variable $Z$. By the displacement theorem \cite{BaccelliBlaszczyszyn} and equivalence theory \cite{Błaszczyszyn}, the density of $\widetilde{\Phi}_{I_m}$ is equal to the density of the original point process $\Phi_{I_m}$ multiplied by the fractional $(2/\alpha)^{th}$ moment of $Z$:
\begin{eqnarray} \label{eq:Representation}
\widetilde{\lambda}_{I_m} &=& \lambda_{I_m} E\left[Z^{2/\alpha}\right] \nonumber \\
&=& \lambda_{I_m} \left(Q_S^{2/\alpha} \frac{\lambda_S}{\lambda_M + \lambda_F} + Q_M^{2/\alpha} \frac{\lambda_M}{\lambda_M + \lambda_S}\right)
\end{eqnarray}
This representation serves as a tool to transform a heterogeneous domain into a homogeneous one. The UL SINR at the associated base station can be calculated as:
\begin{equation}
\begin{split}
	SINR_{U,v} = \frac{P_v h_{x_{v}} \left\|x_{v}\right\|^{-\alpha}}{\sum\limits_{{y_j} \in \widetilde{\Phi}_{I_m} } h_{{y_j}} \left\|y_{j}\right\|^{-\alpha} + \sigma^2},
	\label{UL_SINR}
\end{split}	
\end{equation} 
\noindent where $\sigma^2$ is constant noise power and $v \in \{M,S\}$, depending on the association process.
The spectral efficiency for a given case is calculated with the approach derived in \cite{CoverageRateAndrews}, using the transmit power that corresponds to that particular case. The reader can follow the procedures derived for homogeneous user domain and replace the distance distributions to the serving base station with the ones derived in this paper. The final expression for the spectral efficiency is given by:
\begin{eqnarray}
S_{eff}(\textrm{Case }i) = \int\limits_0^\infty \int\limits_0^\infty {e^{-\pi \widetilde{\lambda}_{I_m} \frac{(e^t-1)^{\frac{2}{\alpha}}}{Q_v} y^2 \int\limits_0^\infty \left(\frac{1}{1+ u^{\alpha/2}} \right) \mathrm{d}{u}}} \times \nonumber \\
{e^{-\frac{(e^t-1) y^\alpha \sigma^2}{Q_v}} f_{v|i}(y) \mathrm{d}{t} \mathrm{d}{y}}%
\end{eqnarray}

The definition of energy efficiency in (\ref{EEdef}) shows that it is closely related to the spectral efficiency. In this sense, the energy efficiency for the devices in Region $i$ is calculated as:
\begin{equation}
E_{eff}(\textrm{Case }i) = \frac{W}{P_{tot}}\textrm{log}_2 {\left( 1+SINR_{U,v}\right)}
\end{equation}
\noindent where $v=M$ for Case 1 and $v=S$ for Cases 2 and 4. The average energy efficiency in the system is evaluated as:
\begin{eqnarray}
		E_{eff} = \sum_{i=1}^4 {E_{eff}(\textrm{Case }i)\Pr(\textrm{Case }i)}
		\label{EEdef_Casei}
\end{eqnarray}

\section{Numerical results}
\label{sec:NumericalResults}

In this section we evaluate numerically the results presented in Section \ref{sec:Analysis}.

\begin{figure}[t!]
	\centering
		\includegraphics[width=8.3cm]{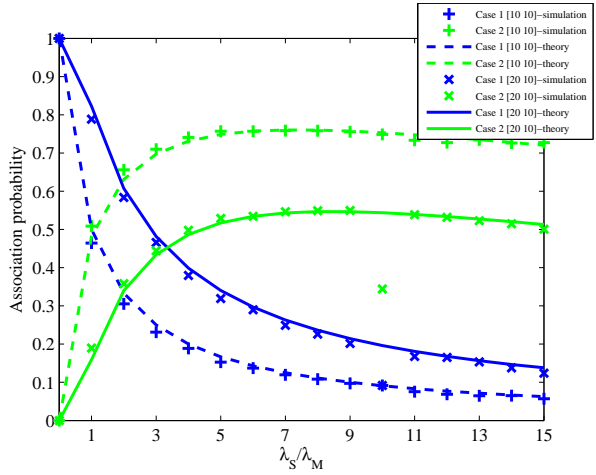}
		\caption{Association probability for different base stations densities ($P_M=46$ dBm, $P_S=20$ dBm, $\alpha$=3).}
		\label{fig:AssocProbBS_density}
\end{figure}

The association regions presented in Fig. \ref{fig:AssocRegions} show that by adapting UL transmit power, one can change the probabilities for association with Cases 1 and 2. Therefore, the results in Fig. \ref{fig:AssocProbBS_density} are focused on these two cases only. The authors in \cite{KaterinaPetarLiljanaDUD} showed that with homogeneous users ($Q_M=Q_S$), increasing SBS density relative to the density of MBSs, the probability for Case 2 initially increases and then starts decreasing at the expense of the probability for Case 4. This trend with Case 2 is also observable here. Additionally, we can manipulate with the probability for Case 2 by changing the uplink power levels relative to each other. Fig. \ref{fig:AssocProbBS_density} shows two different power level settings, $[Q_M, Q_S]=[20, 10]$ dBm and $[Q_M, Q_S]=[10, 10]$ dBm. We can observe that increasing $Q_M$, increases the percentage of devices in Region 1 and lowering $Q_M$ increases the percentage of devices in Region 2, i.e. increases the probability for decoupled access.

\begin{figure}[t!]
	\centering
		\includegraphics[width=8.3cm]{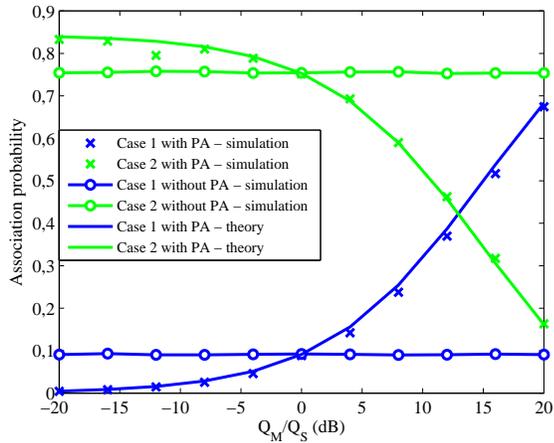}
		\caption{Association probability for different UL transmit powers ($\lambda_S=10\lambda_M$, $P_M=46$ dBm, $P_S=20$ dBm, $\alpha$=3).}
		\label{fig:AssocProbPower}
\end{figure}

Fig. \ref{fig:AssocProbPower} shows the association probability for different UL transmit power levels. It is important to note that the relative UL transmit power is important, not the absolute values of the transmit powers. This can be also concluded by the association probabilities given by (\ref{Pcase1}), (\ref{Pcase2}) and (\ref{Pcase4}), where only the ratio appears. Increasing the ratio $Q_M/Q_S$, the probability for Case 1 increases. The intersection of the curves for Case 1 (with and without power adaptation) and the curves for Case 2 is the point where $Q_M=Q_S$ and the ratio is 0. The intersection between the curves for the Cases 1 and 2 depends on the densities of the base stations. For $\lambda_S<10\lambda_M$, the intersection shifts to the left. It can be concluded that by changing the ratio of the uplink transmit powers, one can achieve soft control of the percentage of devices with decoupled access.

\begin{figure}[t!]
	\centering
		\includegraphics[width=8.3cm]{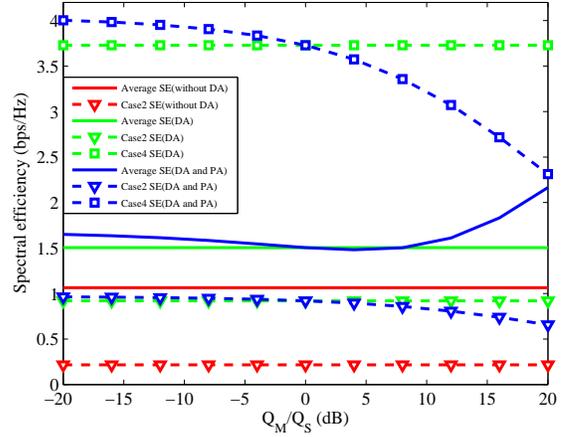}
		\caption{Spectral efficiency (SE) for different UL transmit power and three different settings: (i) Without decoupled access (DA) and without power adaptation (PA), (ii) With DA and without PA and (iii) With DA and with PA ($\lambda_S=10\lambda_M$, $P_M=46$dBm, $P_S=20$dBm).}
		\label{fig:SpectralEff}
\end{figure}

Fig. \ref{fig:SpectralEff} shows the spectral efficiency for three different scenarios: (i) without decoupled access and without power adaptation, (ii) with decoupled access and without power adaptation and (iii) with decoupled access and with power adaptation. The results for the first scenario are obtained by the work conducted in \cite{CoverageRateAndrews}, adapting to the needs of this paper. The results for the second scenario are obtained by taking $Q_M=Q_S$ in the analysis presented in Section III-C. Decoupled DL/UL access by itself is an improvement of the traditional system operation, based on DL received signal power association. Decoupling DL/UL without power adaptation achieves better spectral efficiency for the devices in Case 2, which significantly reflects in the overall spectral efficiency because significant percentage of the devices are in Region 2. The proposed model with two-level UL power adaptation is superior compared to the system with decoupled access and without power adaptation in terms of average spectral efficiency in the system. Basically, by adapting $Q_M$ and $Q_S$, we can make a trade-off between the spectral efficiencies of the separate cases in order to achieve higher overall spectral efficiency in the system. When $Q_M=Q_S$ the average spectral efficiency for the system with decoupled access using power adaptation is equal to the spectral efficiency for the system with decoupled access only.

\begin{figure}[t!]
	\centering
		\includegraphics[width=8.3cm]{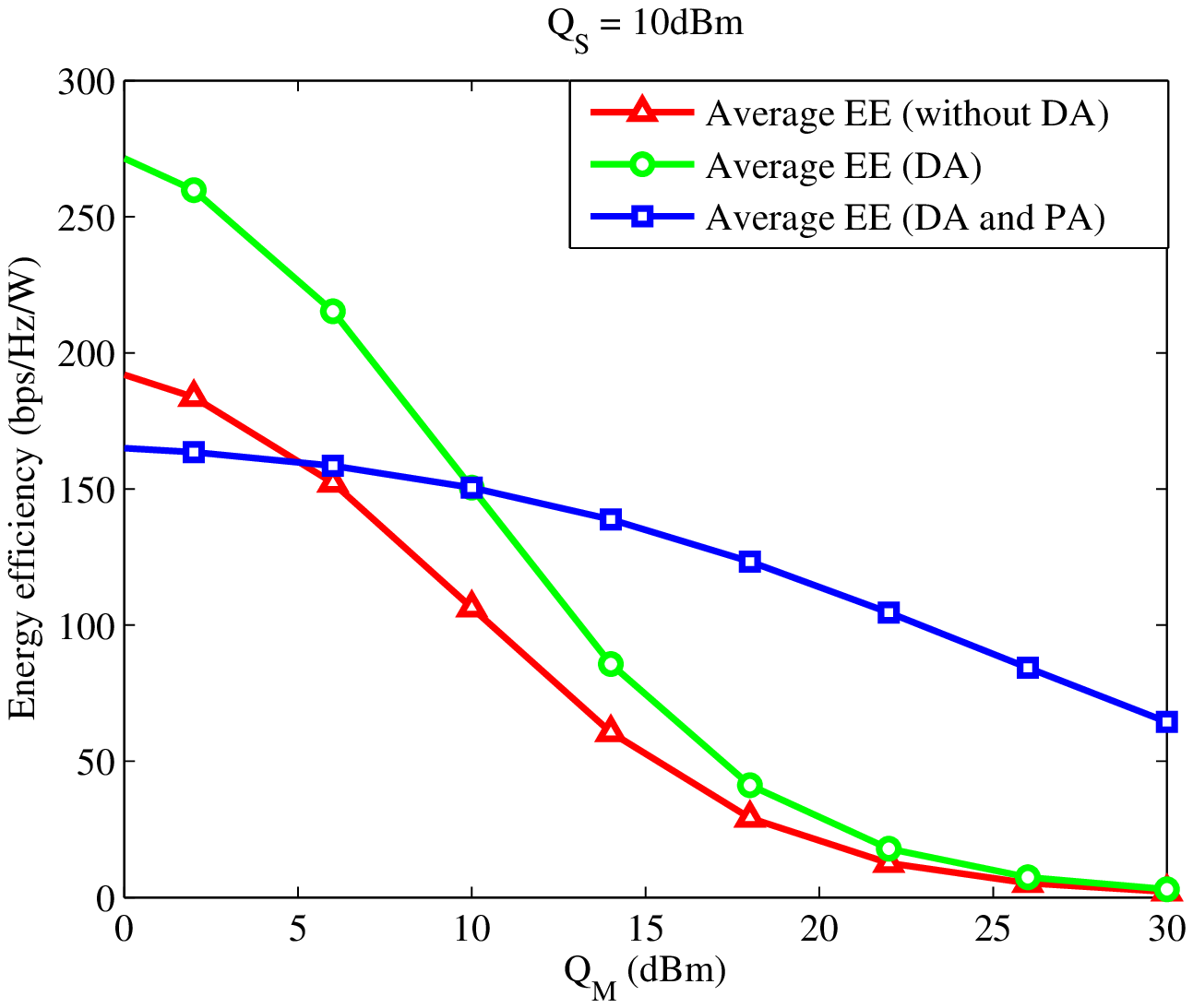}
		\caption{Energy efficiency (EE) for different UL transmit power and three different settings: (i) Without decoupled access (DA) and without power adaptation (PA), (ii) With DA and without PA and (iii) With DA and with PA.}
		\label{fig:EnergyEff_M}
\end{figure}

\begin{figure}[t!]
	\centering
		\includegraphics[width=8.3cm]{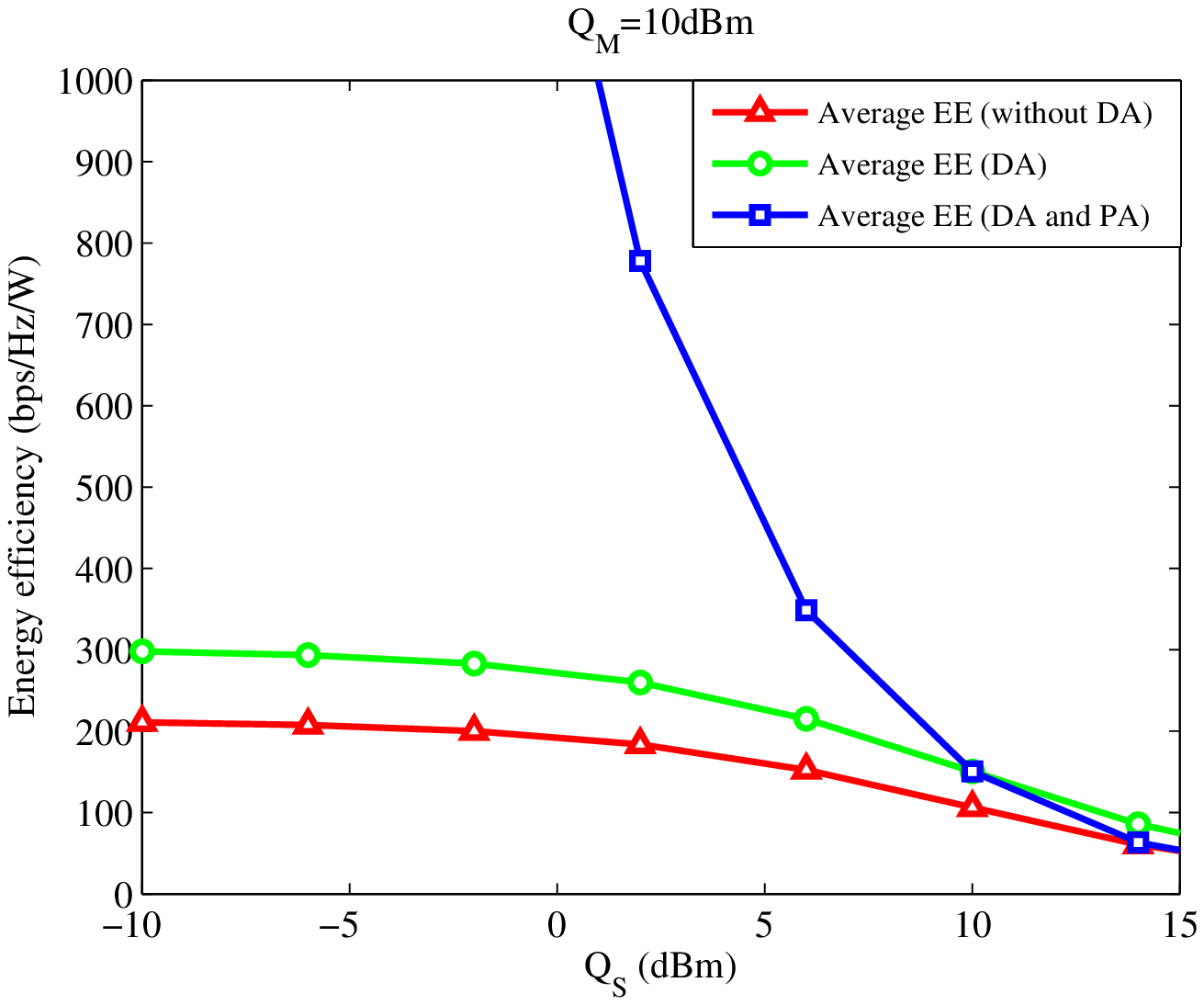}
		\caption{Energy efficiency (EE) for different UL transmit power and three different settings: (i) Without decoupled access (DA) and without power adaptation (PA), (ii) With DA and without PA and (iii) With DA and with PA.}
		\label{fig:EnergyEff_S}
\end{figure}

Fig. \ref{fig:EnergyEff_M} and Fig. \ref{fig:EnergyEff_S} show the average energy efficiency for two different UL power variations. In Fig. \ref{fig:EnergyEff_M}, $Q_S$ is fixed to 10 dBm and $Q_M$ is varying from 0 to 30 dBm. In Fig. \ref{fig:EnergyEff_S}, we fix $Q_M$ to 10 dBm and vary $Q_S$ from -10 dBm to 15 dBm. It can be noted that the spectral efficiency depends only on the ratio between the transmit powers, while the energy efficiency depends on the absolute values of $Q_M$ and $Q_S$. This is due to the fact that the spectral efficiency is mainly influenced by the pdf of the distance to the serving base station, which also depends on the ratio only. On the other hand, calculation of energy efficiency requires normalization by the absolute values of the consumed power, leading to direct dependence on the absolute values rather than on the relative values. 

By comparing Fig. \ref{fig:SpectralEff} with both of the figures for energy efficiency, Fig. \ref{fig:EnergyEff_M} and \ref{fig:EnergyEff_S}, we can see that there is a region for $Q_M>Q_S$, where both, spectral and energy efficiency, are improved. Also it can be noted that the average energy efficiency for the system with decoupled access is always superior compared to the system without decoupling. This is due to the fact that in both systems, the devices transmit with the same transmit power, but with decoupling they achieve more bits per second.

As a final remark, decoupled access is a mechanism that improves both the spectral efficiency and the energy efficiency by preserving the same power and the bandwidth in the system, but changing only the association scheme. 

\section{Conclusion}
\label{conclusion}

In this paper we analyze the performance of a system with decoupled DL/UL access using simple UL power adaptation in a two-tier heterogeneous cellular network. The main results in the paper clearly show that a simple, two-level power adaptation can bring improvement both for the spectral and the energy efficiency in the system. Furthermore, the association and the power adaptation provides some kind of fairness among the devices in order to improve the average performance of the system. 
This initial observations on spectral and energy efficiency with decoupling open new possibilities for joint optimization of UL transmit powers in order to meet certain spectral and energy efficiency targets. 
Our future work will be focused on finding optimal UL transmit power levels in order to meet the requirements for devices in a certain service region.


\section*{Acknowledgment}

Part of this work has been performed in the framework of the FP7 project ICT-317669 METIS, which is partly funded by the European Union. The author would like to acknowledge the contributions of their colleagues in METIS, although the views expressed are those of the authors and do not necessarily represent the project.



%

\end{document}